\documentclass[journal]{IEEEtran}

\usepackage{graphicx}
\usepackage{amssymb}
\usepackage{amsmath}

\usepackage{enumerate}
\usepackage{threeparttable}
\usepackage{multirow}
\usepackage{bm}
\usepackage{soul, color, xcolor}
\usepackage{enumitem}
\usepackage[colorlinks]{hyperref}

\usepackage[ruled,linesnumbered]{algorithm2e}

\SetKwComment{Comment}{// }{}
\usepackage{booktabs}

\usepackage{booktabs}
\usepackage{graphicx}
\usepackage{colortbl}
\usepackage{xcolor}
\usepackage{amssymb} 

\usepackage{multirow}

\begin{document}

\author{{Hongjian Ma}, 
        Wenxin Huang,
        Yan Zhang, 
        Zhifei Li, \textit{Senior Member}, \textit{IEEE}, \\
        and {Zheng Wang, \textit{Senior Member}, \textit{IEEE}}
	
\IEEEcompsocitemizethanks{
\IEEEcompsocthanksitem This work was supported in part by the National Natural Science Foundation of China (No. 62207011, 62301213), the Natural Science Foundation of Hubei Province of China (No. 2025AFB653). (Corresponding author: Zhifei Li)
\IEEEcompsocthanksitem H. Ma, W. Huang, Y. Zhang, and Z. Li are with the School of Computer Science, Hubei University, Wuhan 430062, China, Hubei Key Laboratory of Big Data Intelligent Analysis and Application, Hubei University, Wuhan 430062, China, and also with the Key Laboratory of Intelligent Sensing System and Security (Ministry of Education), Hubei University, Wuhan 430062, China. Z. Wang is with the School of Computer Science, Wuhan University, Wuhan 430072, China.
 }

} 

\title{Modality-Aware Identity Construction and Counterfactual Structure Learning for \\ ID-Free Multimodal Recommendation}

\maketitle

\markboth{IEEE Transactions on Multimedia}
{\MakeLowercase{\textit{Ma et al.}}: MAIL}


\begin{abstract}
Multimodal recommendation has attracted extensive attention by leveraging heterogeneous modality information to alleviate data sparsity and improve recommendation accuracy. Existing methods have attempted to replace ID embeddings with multimodal features and have achieved promising preliminary results. However, these methods still exhibit the following two limitations: (1) the reconstructed ID representations remain relatively static and fail to fully exploit multimodal semantics; and (2) the graph learning process is insufficient in mining latent long-tail semantic relations and is easily affected by popularity bias. To address these issues, we propose a novel method named Modality-Aware Identity Construction and Counterfactual Structure Learning for ID-free Multimodal Recommendation (MAIL). Specifically, we design a modality-aware identity construction module that dynamically modulates positional encodings with multimodal semantics to construct content-aware ID-free identity representations. Then, we propose a counterfactual structure learning paradigm that mines low-exposure semantic neighbors via popularity penalization and alleviates popularity bias. Extensive experiments are conducted on five public Amazon datasets. Experimental results show that MAIL achieves average improvements of 7.81\% in Recall@10 and 12.81\% in NDCG@10 compared with the baseline models.
Our code is available at \url{https://github.com/HubuKG/MAIL}.
\end{abstract}


\section{Introduction}
\IEEEPARstart{R}{ecommender} systems (RS) have emerged as effective techniques for automatically suggesting content that users may be interested in based on their preferences \cite{1-he2016vbpr, 2-wang2022towards, 3-mmrec}. Traditional RS mainly relies on user-item interactions to learn latent representations \cite{4-park2017also, 5-chen2017attentive, 6-xu2018graphcar}.
However, relying solely on user-item interactions makes it difficult to fully characterize user preferences and item properties \cite{7-wei2019mmgcn, 20-lian2026sign}. To alleviate this issue, multimodal recommender systems (MMRS) further incorporate rich item-side content, such as images and textual descriptions, to complement the semantic insufficiency of interaction signals and enhance user and item representations \cite{8-xu2026vi, 9-guo2025self}. Benefiting from the fine-grained characterization of item attributes and user interests provided by multimodal content, MMRS have gradually become an important research direction in recommender systems \cite{10-xu2026survey}.

\begin{figure}[t]
    \centering    
    \includegraphics[width=\columnwidth]{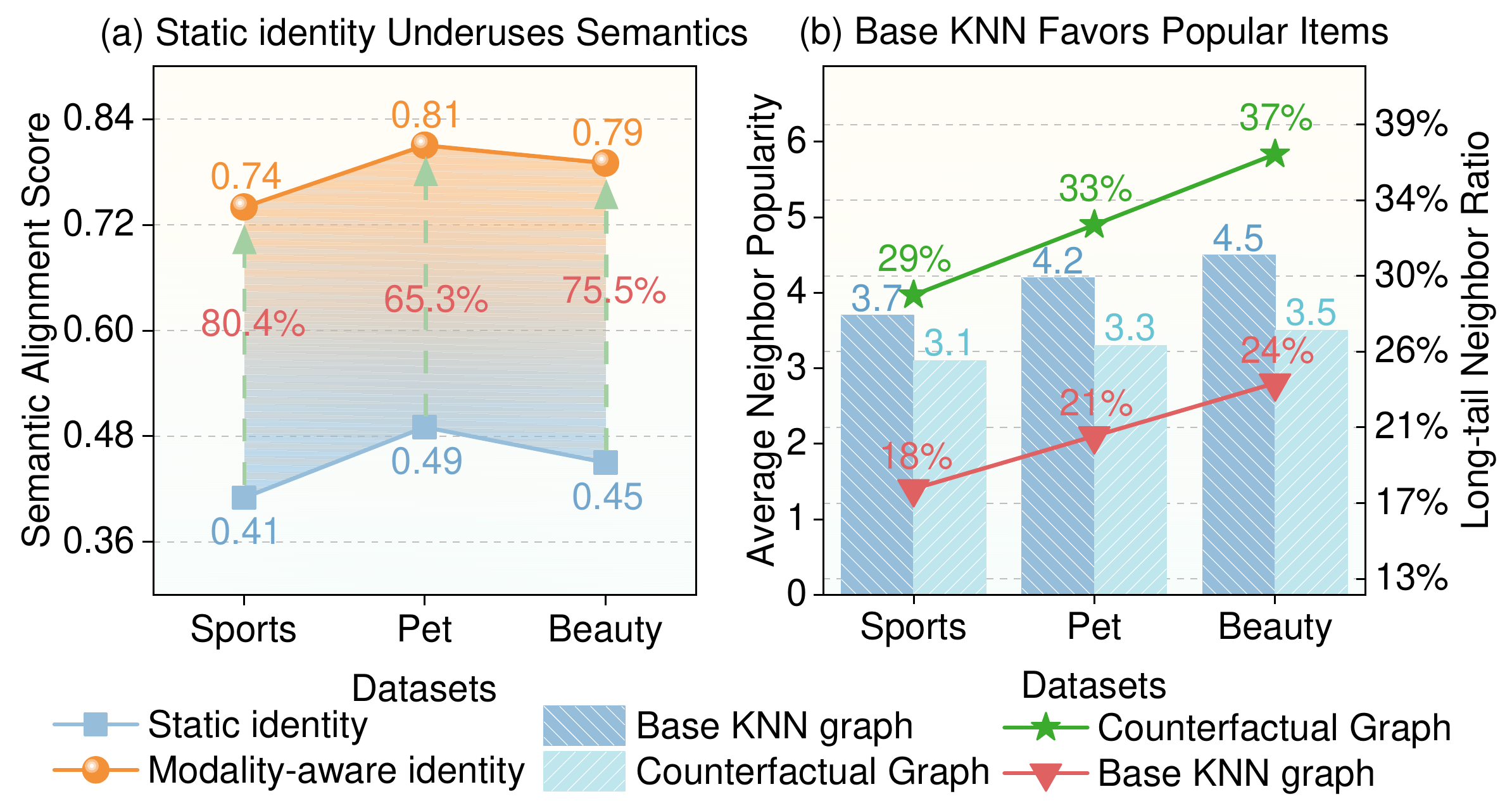}
    \caption{Motivation of MAIL. (a) The static identity construction in existing ID-free methods exhibits weak semantic alignment, making it difficult to fully exploit multimodal information. (b) The base KNN graph favors popular neighbors and provides insufficient long-tail coverage, revealing a clear popularity bias.}
    \label{f1}
\end{figure}

The main challenge for MMRS lies in how to effectively integrate heterogeneous modality information into the collaborative filtering framework \cite{11-yang2025fitmm}. 
Early studies \cite{1-he2016vbpr, 12-tang2019adversarial} typically directly fuse visual or textual features with ID embeddings. Then, graph neural networks (GNNs) have been introduced into multimodal recommendation to model high-order collaborative relations over the user-item interaction graph and fuse multimodal information during graph propagation \cite{13-wang2021dualgnn, 14-zhang2021mining}.
Recently, some studies \cite{15-li2025idfree, 16-zhou2025learning} have further rethought the necessity of ID embeddings in MMRS, pointing out that ID representations can provide node distinguishability but contain limited semantic information and may weaken the model’s generalization ability to unseen users or items. Based on this, ID-free multimodal recommendation methods have begun to directly replace traditional learnable ID embeddings with multimodal features.

Despite the remarkable success of existing methods, there remain two limitations:
\textbf{(1) ID-free methods directly remove ID embeddings and reconstruct identity information with multimodal features, but this process is usually static and fails to fully exploit multimodal semantics}. Existing studies \cite{21-wang2022towards,15-li2025idfree, 22-xu2025mentor} demonstrate that ID embeddings provide node distinguishability but contain limited semantic information and may lead to representation shift. Directly removing ID embeddings also weakens the identity distinguishability of users and items. Therefore, the existing method \cite{15-li2025idfree} uses static positional encodings to replace ID embeddings and supplement node distinguishability. As shown in Fig. \ref{f1} (a), we analyze this issue by calculating the semantic alignment score between identity representations and multimodal semantic anchors. The results show that identity representations constructed by static positional encodings exhibit significantly lower alignment, indicating their limited ability to fully exploit multimodal information. \textbf{(2) The graph learning process of existing GNNs is easily affected by popularity bias, making popular items dominate neighborhood propagation and limiting the mining of low-exposure long-tail semantic relations.} Existing GNN-based methods \cite{17-meng2025tamer, 18-zhou2023bootstrap, 19-wei2023multi} can effectively model high-order collaborative relations and capture latent user interests through the user-item interaction graph. As shown in Fig. \ref{f1} (b), we analyze the neighborhood distribution of the base item-item KNN graph by measuring the average neighbor popularity and the long-tail neighbor ratio. The results show that the base item-item KNN graph tends to select more popular items as neighbors and provides insufficient long-tail neighbor coverage, indicating its limited ability to fully mine low-exposure semantic relations.

To address the above limitations, we propose a novel method named \underline{\textbf{M}}odality-\underline{\textbf{A}}ware \underline{\textbf{I}}dentity Construction and Counterfactual Structure \underline{\textbf{L}}earning for ID-free Multimodal Recommendation (\textbf{MAIL}). Specifically, we design a modality-aware identity construction (MAIC) module that generates modality-aware gates from textual and visual semantics and dynamically modulates positional encodings to transform static identity signals into content-aware ID-free identity representations. Such content-aware identity representations can explicitly incorporate multimodal semantics into the identity construction process, which enhances both node distinguishability and semantic expressiveness. Then, we propose a counterfactual structure learning (CSL) module that constructs low-exposure yet semantically relevant potential neighbors based on counterfactual reasoning \cite{23-wei2021model, 24-zhang2021causal} and integrates them with the basic item-item KNN graph. This enhances the structural expressiveness of the original user-item interaction graph and mitigates the propagation bias dominated by popular items \cite{25-xv2022neutralizing}. Moreover, we design two self-supervised tasks to strengthen counterfactual structure learning from the perspectives of cross-layer structural alignment and decoupled discrimination, improving the retainability and discriminability of latent semantic structures during graph propagation.

To validate the superiority of MAIL, we conduct extensive experiments on five public datasets, including Baby, Sports, Clothing, Pet, and Beauty. Experimental results show that MAIL achieves state-of-the-art performance in terms of Recall@\textit{K} and NDCG@\textit{K}. In addition, we conduct ablation studies to verify the effectiveness of each component in MAIL and evaluate the contribution of different multimodal features used in MAIL to recommendation performance.

In summary, our contributions are as follows:
\begin{itemize}
    \item We design a modality-aware identity construction (MAIC) module, which generates modality-aware gates from modality semantics and dynamically modulates positional encodings to construct content-aware identity representations for ID-free MMRS.
    \item We propose a counterfactual structure learning module that reconstructs item-item semantic neighborhoods from a counterfactual perspective and integrates them into the augmented heterogeneous graph, which helps mine low-exposure yet semantically relevant neighbors and alleviate popular-item-dominated propagation bias.
    \item We conduct extensive experiments on five public benchmark datasets. In terms of Recall@10 and NDCG@10, MAIL achieves average performance improvements of 7.81\% and 12.81\%, respectively.
\end{itemize}

The remainder of this paper is organized as follows. Section \ref{section2} reviews the related work. Section \ref{section3} details the proposed MAIL framework. Section \ref{section4} reports the experimental results and analyses. Section \ref{section5} concludes this paper.

\begin{figure*}[t]
    \centering
     \includegraphics[width=1.0\textwidth]{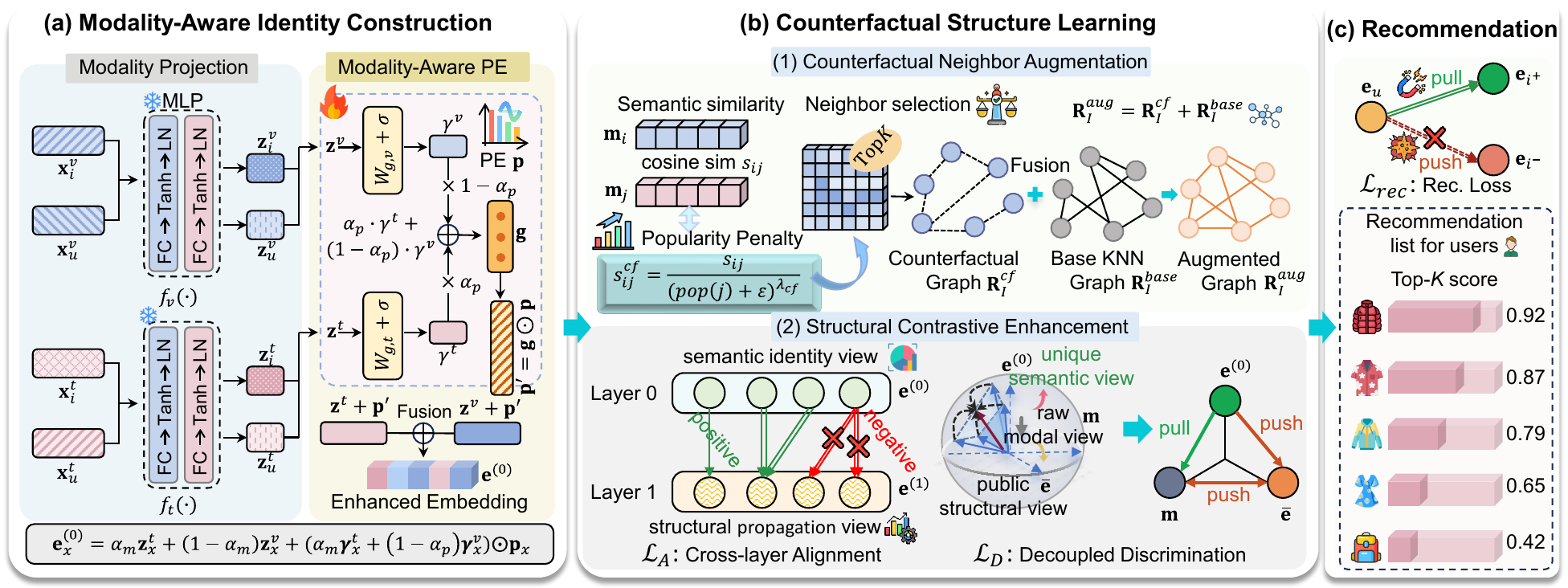}
    \caption{Schematic illustration of our proposed MAIL. (a) MAIC module, which dynamically modulates positional encodings to construct content-aware ID-free identity representations. (b) CSL module, which mines low-exposure semantic neighbors and alleviates popular-item-dominated propagation bias. (c) Recommendation module, which calculates user-item preference scores and optimizes Top-\textit{K} recommendation results.}
    \label{f2}
\end{figure*}

\section{Related work}
\label{section2}

\subsection{ID-based Multimodal Recommendation}
ID-based multimodal recommendation (MMRS) refers to recommendation methods that jointly utilize user/item ID embeddings and multimodal features such as images and text within the collaborative filtering framework to model collaborative signals and semantic information \cite{39-ma2026let, 29-ong2025spectrum}. BPR \cite{26-rendle2009bpr} and LightGCN \cite{27-he2020lightgcn} are two representative methods in traditional collaborative filtering. VBPR \cite{1-he2016vbpr} first incorporates multimodal information into collaborative filtering by integrating visual features with ID embeddings. LGMRec \cite{28-guo2024lgmrec} further models both local and global user interests via graph and hypergraph learning. Moreover, MENTOR \cite{22-xu2025mentor} designs multi-level self-supervised tasks to align different modalities while preserving historical interaction information. SMORE \cite{29-ong2025spectrum} and FITMM \cite{11-yang2025fitmm} further emphasize frequency-domain modeling by decomposing multimodal signals into spectral components to suppress modality-specific noise and enable more adaptive modality fusion. The latest SiMGR \cite{20-lian2026sign} further emphasizes signed graph modeling by explicitly distinguishing positive and negative feedback and integrating them with multimodal features for more precise preference learning. Different from the above ID-based methods, our MAIL further moves toward ID-free MMRS by fully exploiting multimodal semantics for identity construction and structure learning to reduce the reliance on learnable ID embeddings.

\subsection{ID-free Multimodal Recommendation}
ID-free multimodal recommendation \cite{15-li2025idfree, 33-hu2023adaptive} aims to remove traditional learnable ID embeddings and instead relies on multimodal content features such as text and images to construct user and item representations. ID-free multimodal recommendation is still in its infancy, and existing studies \cite{31-wang2025generative, 32-penha2025semantic, 34-zhang2024ninerec} mainly explore how to replace ID representations with multimodal semantics in other recommendation scenarios such as generative recommendation and sequential recommendation. In particular, the representative method IDFREE \cite{15-li2025idfree} uses item multimodal features as initial representations and introduces static positional encodings to supplement the unique identity distinguishability of users and items. However, the positional encodings in IDFREE \cite{15-li2025idfree} remain static and cannot dynamically adjust identity signals according to multimodal semantics. In contrast, our MAIL enhances the content awareness of ID-free identity representations and the long-tail semantic mining ability of graph structures through modality-aware identity construction and counterfactual structure learning, respectively.

\subsection{Counterfactual Learning}
Observed interactions in RS are easily affected by exposure and popularity bias, while counterfactual learning simulates recommendation scenarios where such biases are removed \cite{35-wei2021model}. PPAC \cite{36-ning2024debiasing} proposes a personal popularity-aware counterfactual framework that incorporates users’ personalized popularity preferences into counterfactual modeling. Moreover, CoDeR \cite{37-tang2025coder} starts from evolving user demands in sequential recommendation and uses counterfactual demand reasoning to identify and distinguish genuine demand shifts from noisy behaviors. In addition, SR-GCA \cite{38-huang2025social} generates counterfactual social edges and item edges through graph-level counterfactual augmentation.
Therefore, our MAIL further analyzes the popular-neighbor dominance problem in multimodal graph learning and proposes a counterfactual structure learning paradigm to mine low-exposure semantic neighbors and alleviate popularity bias.

\section{Methodology}
\label{section3}
In this section, we present the proposed MAIL in detail.
As shown in Fig. \ref{f2}, MAIL mainly consists of three components. (a) Modality-Aware Identity Construction constructs content-aware ID-free identity representations by dynamically modulating positional encodings. (b) Counterfactual Structure Learning mines low-exposure semantic neighbors and alleviates popular-item-dominated propagation bias. (c) Recommendation calculates preference scores based on the final user and item representations.
\subsection{Preliminaries}
In MMRS, let $\mathcal{U}$ and $\mathcal{I}$ denote the user set and item set, with $|\mathcal{U}|$ users and $|\mathcal{I}|$ items, respectively. The user-item interaction matrix is denoted as $\mathbf{R}\in\{0,1\}^{|\mathcal{U}|\times|\mathcal{I}|}$, where $\mathbf{R}_{u,i}=1$ indicates that user $u$ has interacted with item $i$, and $\mathbf{R}_{u,i}=0$ otherwise. This paper considers textual and visual modalities, denoted as $\mathcal{M}=\{t,v\}$. Based on $\mathbf{R}$, we construct a user-item bipartite graph $\mathcal{G}=(\mathcal{U},\mathcal{I},\mathcal{E})$, where $\mathcal{E}$ denotes the observed interaction edges. The goal of MAIL is to learn user representation $\mathbf{e}_u$ and item representation $\mathbf{e}_i$ from multimodal features and graph structures without learnable ID embeddings, and predict the preference score $\hat{r}_{u,i}$ for Top-$K$ recommendation.

\subsection{Modality-Aware Identity Construction}
\subsubsection{Modality Projection}
To map textual and visual features into a unified latent space, MAIL first projects modality features of users and items. For each node $x\in\mathcal{U}\cup\mathcal{I}$ and modality $m\in\{t,v\}$, we have:
\begin{equation}
\mathbf{z}_x^m=\mathrm{LN}\left(\tanh\left(\mathbf{W}_m\mathbf{x}_x^m+\mathbf{b}_m\right)\right),
\end{equation}
where $\mathbf{x}_x^m$ is the raw feature of node $x$ under modality $m$, $\mathbf{z}_x^m\in\mathbb{R}^d$ is the projected modality representation, $\mathbf{W}_m$ and $\mathbf{b}_m$ are modality-specific learnable parameters, and $\mathrm{LN}(\cdot)$ denotes Layer Normalization. Since users usually have no explicit multimodal content, their modality features are obtained by averaging the features of interacted items:
\begin{equation}
    \mathbf{x}_u^m=\frac{1}{|\mathcal{N}_u|}\sum_{i\in\mathcal{N}_u}\mathbf{x}_i^m,
\end{equation}
where $\mathcal{N}_u$ denotes the historical interaction item set of user $u$, and $\mathbf{x}_i^m$ denotes the raw feature of item $i$ under modality $m$.
\subsubsection{Modality-Aware PE}
Traditional ID-free methods usually add fixed positional encodings to modality representations. However, this strategy cannot dynamically adjust identity signals according to multimodal semantics. To this end, MAIL generates modality-aware gates from textual and visual semantics:
\begin{equation}
    \boldsymbol{\gamma}_x^t=\sigma(\mathbf{W}_g^t\mathbf{z}_x^t+\mathbf{b}_g^t),\quad
\boldsymbol{\gamma}_x^v=\sigma(\mathbf{W}_g^v\mathbf{z}_x^v+\mathbf{b}_g^v),
\end{equation}
where $\boldsymbol{\gamma}_x^t$ and $\boldsymbol{\gamma}_x^v$ denote the positional encoding gates generated from textual and visual modalities, $\sigma(\cdot)$ is the Sigmoid function, and $\mathbf{W}_g^t,\mathbf{W}_g^v$ are learnable parameters of the gating networks.

Then, MAIL fuses the two modality-aware gates and dynamically modulates the base positional encoding:
\begin{equation}
    \mathbf{g}_x=\alpha_p\boldsymbol{\gamma}_x^t+(1-\alpha_p)\boldsymbol{\gamma}_x^v,\quad
\mathbf{p}_x'=\mathbf{g}_x\odot\mathbf{p}_x,
\end{equation}
where $\mathbf{p}_x$ is the original positional encoding of node $x$, $\mathbf{g}_x$ is the modality-aware gate, $\alpha_p$ controls the fusion ratio of textual and visual gates, $\odot$ denotes element-wise multiplication, and $\mathbf{p}_x'$ denotes the modulated modality-aware positional encoding.

Finally, the modulated positional encoding is injected into textual and visual representations. The two modalities are then fused to obtain the content-aware ID-free identity representation:
\begin{equation}
    \mathbf{e}_x^{(0)}
=
\alpha_m(\mathbf{z}_x^t+\mathbf{p}_x')
+
(1-\alpha_m)(\mathbf{z}_x^v+\mathbf{p}_x'),
\end{equation}
where $\mathbf{e}_x^{(0)}$ denotes the initial ID-free identity representation of node $x$, and $\alpha_m$ is the fusion weight between textual and visual modalities. This representation explicitly incorporates multimodal semantics into identity construction, making positional encoding a content-aware identity signal dynamically modulated by textual and visual content rather than a static identity signal.

\subsection{Counterfactual Structure Learning}
\subsubsection{Counterfactual Neighbor Augmentation}
To alleviate the tendency of the base item-item KNN graph toward popular items, MAIL constructs counterfactual neighbors based on multimodal semantic similarity and popularity penalization. First, textual and visual item representations are fused into a semantic representation:
\begin{equation}
    \mathbf{h}_i=\alpha_m\mathbf{z}_i^t+(1-\alpha_m)\mathbf{z}_i^v ,
\end{equation}
where $\mathbf{z}_i^t$ and $\mathbf{z}_i^v$ denote the textual and visual representations of item $i$, $\alpha_m$ is the modality fusion weight, and $\mathbf{h}_i$ denotes the fused item semantic representation.

Then, we calculate the semantic similarity between item $i$ and item $j$:
\begin{equation}
    s_{ij}=\frac{\mathbf{h}_i^\top \mathbf{h}_j}{\|\mathbf{h}_i\|_2\|\mathbf{h}_j\|_2},
\end{equation}
where $s_{ij}$ denotes the cosine similarity of item pair $(i,j)$.

To reduce the dominance of popular items in neighbor selection, MAIL introduces a popularity penalty for candidate neighbor $j$:
\begin{equation}
    s_{ij}^{cf}=\frac{s_{ij}}{(\mathrm{pop}(j)+\epsilon)^{\lambda_{cf}}},
\end{equation}
where $\mathrm{pop}(j)=\log(1+n_j)$ denotes the popularity of item $j$, $n_j$ is the number of interactions of item $j$, $\epsilon$ is a numerical stability term, $\lambda_{cf}$ controls the penalty strength, and $s_{ij}^{cf}$ denotes the counterfactual neighbor selection score.

Based on $s_{ij}^{cf}$, MAIL selects Top-$K_{cf}$ low-exposure yet semantically relevant neighbors and constructs the counterfactual item-item graph:
\begin{equation}
    \mathcal{N}_{i}^{cf}=\operatorname{TopK}_{j\in\mathcal{I},j\neq i}(s_{ij}^{cf}),\quad
\mathbf{R}_{I}^{cf}(i,j)=s_{ij},\ j\in\mathcal{N}_{i}^{cf},
\end{equation}
where $\mathcal{N}_{i}^{cf}$ denotes the counterfactual neighbor set of item $i$, and $\mathbf{R}_{I}^{cf}$ denotes the counterfactual item-item graph. The edge weight adopts the original semantic similarity $s_{ij}$ to preserve true semantic relevance.

Finally, the counterfactual item-item graph is fused with the base item-item KNN graph to obtain the enhanced item-item graph:
\begin{equation}
\mathbf{R}_I^{aug} = \mathbf{R}_I^{base} + \eta \mathbf{R}_I^{cf},
\end{equation}
where $\mathbf{R}_I^{base}$ denotes the original item-item KNN graph
constructed from multimodal item representations, $\eta$ is the
fusion weight of counterfactual edges, and $\mathbf{R}_I^{aug}$ denotes the
enhanced item-item graph. Since users have no explicit multimodal
content, we obtain user representations by averaging the features of
their interacted items and construct the user-user KNN graph
$\mathbf{R}_U$ accordingly. Combined with $\mathbf{R}_U$ and the
user-item interaction graph, the enhanced graph is constructed as:
\begin{equation}
    \mathbf{A}^{aug}=
\begin{bmatrix}
\mathbf{R}_{U} & \mathbf{R} \\
\mathbf{R}^{\top} & \mathbf{R}_{I}^{aug}
\end{bmatrix},
\end{equation}
where $\mathbf{R}_U$ denotes the user-user KNN graph,
$\mathbf{R}$ denotes the original user-item interaction matrix,
and $\mathbf{A}^{aug}$ is used for subsequent graph propagation.
\subsubsection{Structural Contrastive Enhancement}
After obtaining the enhanced graph $\mathbf{A}^{aug}$, MAIL adopts LightGCN \cite{27-he2020lightgcn} for structural propagation:
\begin{equation}
    \mathbf{E}^{(l+1)}=\widehat{\mathbf{A}}^{aug}\mathbf{E}^{(l)},
\quad
\widehat{\mathbf{A}}^{aug}=\mathbf{D}^{-\frac{1}{2}}\mathbf{A}^{aug}\mathbf{D}^{-\frac{1}{2}},
\end{equation}
where $\mathbf{E}^{(l)}$ denotes the user and item representations at the $l$-th layer, $\widehat{\mathbf{A}}^{aug}$ is the normalized enhanced adjacency matrix, and $\mathbf{D}$ is the degree matrix.

To preserve counterfactual structures during propagation, MAIL introduces a cross-stage alignment loss that aligns pre-propagation semantic identity representations with post-propagation structural representations. For each positive pair $(u,i)$, we define:
\begin{equation}
    \mathcal{L}_{A}^{u}
=
-\log
\frac{
\sum_{i\in\mathcal{P}(u)}\exp(\mathrm{sim}(\mathbf{e}_{u}^{(0)},\mathbf{e}_{i}^{(1)})/\tau_A)
}{
\sum_{j\in\mathcal{B}_I}\exp(\mathrm{sim}(\mathbf{e}_{u}^{(0)},\mathbf{e}_{j}^{(1)})/\tau_A)
},
\end{equation}
where $\mathbf{e}_{u}^{(0)}$ denotes the initial semantic identity representation of user $u$, $\mathbf{e}_{i}^{(1)}$ denotes the structural representation of item $i$ after one-layer graph propagation, $\mathcal{P}(u)$ is the positive item set of user $u$, $\mathcal{B}_I$ is the candidate item set in the batch, and $\tau_A$ is the temperature parameter. Similarly, we obtain the item-side alignment loss $\mathcal{L}_{A}^{i}$. The overall alignment loss is:
$$
\mathcal{L}_{A}=\mathcal{L}_{A}^{u}+\mathcal{L}_{A}^{i}.
$$

Moreover, MAIL designs a decoupled discrimination loss to preserve raw multimodal semantics in the initial identity view and distinguish it from the public structural view after graph propagation:
\begin{equation}
    \mathcal{L}_{D}^{i}
=
-\log
\frac{
\exp(\mathrm{sim}(\mathbf{e}_{i}^{(0)},\mathbf{m}_{i})/\tau_D)
}{
\sum_{j\in\mathcal{B}_I}\exp(\mathrm{sim}(\mathbf{e}_{i}^{(0)},\bar{\mathbf{e}}_{j})/\tau_D)
},
\end{equation}
where $\mathbf{e}_{i}^{(0)}$ denotes the initial semantic identity representation of item $i$, $\mathbf{m}_{i}$ denotes the raw multimodal fused representation of item $i$, $\bar{\mathbf{e}}_{j}$ denotes the final graph-propagated representation of item $j$, and $\tau_D$ is the temperature parameter. This loss encourages the initial identity representation to approach raw multimodal semantics while moving away from over-fused public structural representations.

Similarly, we obtain the user-side decoupled discrimination loss $\mathcal{L}_{D}^{u}$. The overall discrimination loss is:
\begin{equation}
    \mathcal{L}_{D}=\mathcal{L}_{D}^{u}+\mathcal{L}_{D}^{i}.
\end{equation}
Finally, the structural contrastive enhancement objective is defined as:
\begin{equation}
    \mathcal{L}_{SCE}=\mathcal{L}_{A}+\mathcal{L}_{D},
\end{equation}
Through $\mathcal{L}_{A}$ and $\mathcal{L}_{D}$, MAIL strengthens counterfactual structure learning from cross-layer structural preservation and semantic-structural decoupling, making low-exposure semantic neighbors more stable and discriminative during graph propagation.

\subsection{Recommendation}
After graph propagation, MAIL obtains the final user representation $\mathbf{e}_u$ and item representation $\mathbf{e}_i$. During prediction, the preference score of user $u$ on item $i$ is calculated by inner product:
\begin{equation}
    \hat{r}_{u,i}=\mathbf{e}_u^\top \mathbf{e}_i ,
\end{equation}
where $\mathbf{e}_u$ denotes the final representation of user $u$, $\mathbf{e}_i$ denotes the final representation of item $i$, and $\hat{r}_{u,i}$ denotes the predicted preference score. All candidate items are ranked according to $\hat{r}_{u,i}$ to obtain the Top-$K$ recommendation results.

During training, MAIL does not adopt BPR \cite{26-rendle2009bpr} loss. Instead, it uses a softmax ranking loss based on cosine similarity. For a training sample $(u,i^+,\mathcal{I}_u^-)$, the recommendation loss is defined as:
\begin{equation}
\begin{aligned}
\mathcal{L}_{rec}
=
\frac{1}{|\mathcal{B}|}
\sum_{(u,i^+)\in\mathcal{B}}
\log
\sum_{i^-\in\mathcal{I}_u^-}
\exp \bigg(
&\frac{
\mathrm{sim}(\mathbf{e}_u,\mathbf{e}_{i^-})
}{\tau}  \\
&-
\frac{
\mathrm{sim}(\mathbf{e}_u,\mathbf{e}_{i^+})
}{\tau}
\bigg).
\end{aligned}
\end{equation}
where $\mathcal{B}$ denotes the training batch, $i^+$ is the positive item of user $u$, and $\mathcal{I}_u^-$ denotes the negative item set. This loss encourages the user representation to be closer to positive items and farther from negative items.

In addition, MAIL retains the cross-modal InfoNCE constraint in IDFREE \cite{15-li2025idfree} to align textual and visual representations and enhance inter-modal consistency.

The final training objective is:
\begin{equation}
    \mathcal{L}
=
\mathcal{L}_{rec}
+
\lambda_m \mathcal{L}_{m}
+
\lambda_S \mathcal{L}_{SCE},
\end{equation}
where $\mathcal{L}_{rec}$ is the recommendation loss, $\mathcal{L}_{m}$ is the cross-modal alignment loss, $\mathcal{L}_{SCE}$ is the structural contrastive enhancement loss, and $\lambda_m$ and $\lambda_S$ control the weights of the corresponding loss terms.

\begin{algorithm}[t]
\caption{The overall training process of MAIL.}
\footnotesize
\label{alg:mail}
\KwIn{Number of GNN layers $L$, batches $B$, epochs $E$, user and item sets $\mathcal{U}$ and $\mathcal{I}$, user-item interaction matrix $\mathbf{R}$, raw multimodal features $\{\mathbf{x}_i^t\}_{i=1}^{|\mathcal{I}|}$ and $\{\mathbf{x}_i^v\}_{i=1}^{|\mathcal{I}|}$, base positional encodings $\{\mathbf{p}_x\}_{x \in \mathcal{U} \cup \mathcal{I}}$.}
\KwOut{Learned model parameters of MAIL for predicting the preference score $\hat{r}_{u,i}$.}
Initialize model parameters\;
$\mathbf{z}_x^t, \mathbf{z}_x^v \leftarrow \text{Eq. (1)}$ \tcp*[r]{Modality projection}
$\mathbf{z}_u^t, \mathbf{z}_u^v \leftarrow \text{Eq. (2)}$ \tcp*[r]{User features}
$\mathbf{h}_i \leftarrow \text{Eq. (6)}$ \tcp*[r]{Item semantic rep.}
$s_{ij}^{cf} \leftarrow \text{Eqs. (7-8)}$ \tcp*[r]{Popularity penalty}
$\mathbf{R}_I^{cf}, \mathbf{A}^{aug} \leftarrow \text{Eqs. (9-11)}$ \tcp*[r]{Enhanced graph}
\For{$e = 1, \dots, E$}{
    \For{$b = 1, \dots, B$}{
        $\boldsymbol{\gamma}_x^t, \boldsymbol{\gamma}_x^v \leftarrow \text{Eq. (3)}$ \tcp*[r]{Modality-aware gates}
        $\mathbf{g}_x, \mathbf{p}_x^\prime \leftarrow \text{Eq. (4)}$ \tcp*[r]{Modulated PE}
        $\mathbf{e}_x^{(0)} \leftarrow \text{Eq. (5)}$ \tcp*[r]{Initial ID-free rep.}
        \For{$l = 0, \dots, L-1$}{
            $\mathbf{E}^{(l+1)} \leftarrow \text{Eq. (12)}$ \tcp*[r]{Propagation}
        }
        $\mathcal{L}_{A} \leftarrow \text{Eq. (13)}$ \tcp*[r]{Cross-layer alignment}
        $\mathcal{L}_{D} \leftarrow \text{Eqs. (14-15)}$;
        \\
        $\mathcal{L}_{SCE} \leftarrow \text{Eq. (16)}$ \tcp*[r]{Structural loss}
        $\mathcal{L}_{rec} \leftarrow \text{Eq. (18)}$ \tcp*[r]{Recommendation loss}
        $\mathcal{L} \leftarrow \text{Eq. (19)}$ \tcp*[r]{Final loss of MAIL}
        Update model parameters by minimizing $\mathcal{L}$.
    }
}
\end{algorithm}

\subsection{Model Analysis}
\subsubsection{Training Process}
The training procedure of MAIL is shown in Algorithm \ref{alg:mail}. Specifically, MAIL first projects textual and visual features into a unified latent space and generates modality-aware gates from modality semantics to dynamically modulate positional encodings, producing the initial ID-free identity representation $\mathbf{e}_x^{(0)}$. Then, MAIL constructs counterfactual neighbors based on multimodal semantic similarity and popularity penalization, and fuses the counterfactual item-item graph with the base item-item KNN graph to obtain the enhanced item-item graph, which is further integrated with the user-user KNN graph and the user-item interaction graph to form the augmented heterogeneous graph $A^{aug}$. Based on $\mathbf{A}^{aug}$, MAIL performs graph propagation with LightGCN to obtain the final user representation $\mathbf{e}_u$ and item representation $\mathbf{e}_i$. Finally, the model parameters are updated by minimizing the overall loss $\mathcal{L}$.
\subsubsection{Time Complexity of MAIL}
The main computational cost of MAIL comes from the MAIC and CSL modules. For MAIC, the time complexity of modality projection and gate generation is $O((|\mathcal{U}|+|\mathcal{I}|)d^2)$, where $d$ is the hidden dimension. For Counterfactual Neighbor Augmentation, the complexity of item semantic similarity computation is $O(|\mathcal{I}|^2d)$, and the complexity of Top-$K_{cf}$ neighbor selection is $O(|\mathcal{I}|^2)$. The counterfactual graph is pre-constructed only once, so it does not significantly increase the training cost of each epoch. For graph propagation, MAIL adopts LightGCN on the enhanced graph $\mathbf{A}^{aug}$, with a time complexity of $O(L|\mathcal{E}^{aug}|d)$, where $L$ is the number of GNN layers and $|\mathcal{E}^{aug}|$ is the number of edges in the enhanced graph. Structural contrastive enhancement is mainly computed within each batch, with a complexity of $O(|\mathcal{B}|^2d)$. Therefore, the main training complexity of MAIL is $O(L|\mathcal{E}^{aug}|d+|\mathcal{B}|^2d)$, which remains in the same order as mainstream GNN-based MMRS.
\begin{table}[t]
\centering
\caption{Statistics of the Used Datasets.}
\label{t-1}
\begin{tabular}{lrrrr}
\toprule
Dataset & Users & Items & Interactions & Sparsity \\
\midrule
Baby     & 19,445  & 7,050  & 160,792   & 99.88\% \\
Sports   & 35,598  & 18,357 & 296,337   & 99.95\% \\
Clothing & 39,387  & 23,033 & 278,677   & 99.97\% \\
Pet      & 19,856  & 8,510  & 157,836   & 99.91\% \\
Beauty   & 22,363  & 12,101 & 198,502   & 99.93\% \\
\bottomrule
\end{tabular}
\end{table}

\begin{table*}[t]
\centering
\Large
\caption{Comparison of the performance between baseline methods and MAIL method based on Recall@10 and NDCG@10. Best and second-best results are highlighted in \textbf{bold} and \underline{underlined}, respectively. $\uparrow$ indicates a relative improvement of MAIL over the best baseline.}
\label{t-2}
\resizebox{\textwidth}{!}{
\begin{tabular}{l|cc|cc|cc|cc|cc}
\toprule
Dataset & \multicolumn{2}{c|}{Baby} & \multicolumn{2}{c|}{Sports} & \multicolumn{2}{c|}{Clothing} & \multicolumn{2}{c|}{Pet} & \multicolumn{2}{c}{Beauty} \\
\midrule
Metrics & Recall@10 & NDCG@10 & Recall@10 & NDCG@10 & Recall@10 & NDCG@10 & Recall@10 & NDCG@10 & Recall@10 & NDCG@10 \\
\midrule
\rowcolor{gray!20} \multicolumn{11}{c}{Traditional recommendation models} \\
MF-BPR    & 0.0357 & 0.0192 & 0.0432 & 0.0241 & 0.0187 & 0.0279 & 0.0348 & 0.0192 & 0.0301 & 0.0167 \\
LightGCN  & 0.0479 & 0.0257 & 0.0569 & 0.0311 & 0.0340 & 0.0226 & 0.0495 & 0.0295 & 0.0428 & 0.0257 \\
DimCL     & 0.0532 & 0.0286 & 0.0606 & 0.0332 & 0.0398 & 0.0221 & 0.0548 & 0.0226 & 0.0474 & 0.0197 \\
\midrule
\rowcolor{gray!20} \multicolumn{11}{c}{ID-based multimodal recommendation models} \\
LGMRec    & 0.0639 & 0.0337 & 0.0719 & 0.0387 & 0.0555 & 0.0302 & 0.1057 & 0.0584 & 0.0861 & 0.0507 \\
SMORE     & 0.0680 & 0.0365 & 0.0762 & 0.0408 & 0.0659 & 0.0360 & 0.1134 & 0.0626 & 0.0994 & 0.0558 \\
PGL       & 0.0676 & 0.0354 & 0.0789 & 0.0428 & 0.0712 & 0.0385 & 0.1175 & 0.0644 & 0.1031 & 0.0575 \\
COHESION  & 0.0680 & 0.0354 & 0.0752 & 0.0409 & 0.0665 & 0.0358 & 0.1132 & 0.0619 & 0.0992 & 0.0552 \\
SSR       & 0.0728 & 0.0395 & 0.0825 & 0.0449 & 0.0708 & 0.0386 & 0.1221 & 0.0679 & 0.1071 & 0.0606 \\
SRGFormer & 0.0681 & 0.0369 & 0.0758 & 0.0412 & 0.0596 & 0.0330 & 0.1099 & 0.0613 & 0.0963 & 0.0547 \\
LETTER    & 0.0693 & 0.0369 & 0.0811 & 0.0444 & 0.0706 & 0.0386 & 0.1193 & 0.0662 & 0.1045 & 0.0590 \\
SiMGR     & 0.0702 & 0.0382 & 0.0782 & 0.0428 & 0.0712 & 0.0391 & 0.1185 & 0.0663 & 0.1039 & 0.0591 \\
DuGRec    & 0.0734 & 0.0402 & 0.0827 & 0.0462 & 0.0727 & 0.0397 & 0.1235 & 0.0696 & 0.1082 & 0.0621 \\
\midrule
\rowcolor{gray!20} \multicolumn{11}{c}{ID-free multimodal recommendation models} \\
IDFREE    & \underline{0.1147} & \underline{0.0772} & \underline{0.1173} & \underline{0.0770} & \underline{0.1053} & \underline{0.0776} & \underline{0.1354} & \underline{0.0876} & \underline{0.1262} & \underline{0.0797} \\
MAIL (Ours) & \textbf{0.1305} & \textbf{0.0956} & \textbf{0.1226} & \textbf{0.0845} & \textbf{0.1106} & \textbf{0.0828} & \textbf{0.1451} & \textbf{0.0948} & \textbf{0.1370} & \textbf{0.0921} \\
\midrule
\rowcolor{gray!20}
Improvement & 13.78\%$\uparrow$ & 23.83\%$\uparrow$ & 4.52\%$\uparrow$ & 9.74\%$\uparrow$ & 5.03\%$\uparrow$ & 6.70\%$\uparrow$ & 7.16\%$\uparrow$ & 8.22\%$\uparrow$ & 8.56\%$\uparrow$ & 15.56\%$\uparrow$ \\
\bottomrule
\end{tabular}
}
\end{table*}

\begin{table*}[t]
\centering
\LARGE
\caption{Experimental results of module ablation in MAIL. The modules in MAIL all have a significant impact on performance. \textcolor{green!70!black}{$\checkmark$} indicates the use of the module. \textcolor{red}{$\times$} indicates that the module is not used.}
\label{t-3}
\resizebox{\textwidth}{!}{
\begin{tabular}{cccc|cccc|cccc|cccc}
\toprule
\multirow{2}{*}{MAIC} & \multirow{2}{*}{CNA} & \multirow{2}{*}{SCE} & \multirow{2}{*}{MM} & \multicolumn{4}{c|}{Sports} & \multicolumn{4}{c|}{Pet} & \multicolumn{4}{c}{Beauty} \\
\cmidrule(lr){5-8} \cmidrule(lr){9-12} \cmidrule(lr){13-16}
& & & & Recall@10 & Recall@20 & NDCG@10 & NDCG@20 & Recall@10 & Recall@20 & NDCG@10 & NDCG@20 & Recall@10 & Recall@20 & NDCG@10 & NDCG@20 \\
\midrule
\textcolor{green!70!black}{$\checkmark$} & \textcolor{green!70!black}{$\checkmark$} & \textcolor{green!70!black}{$\checkmark$} & \textcolor{green!70!black}{$\checkmark$} & \textbf{0.1226} & \textbf{0.1539} & \textbf{0.0845} & \textbf{0.0926} & \textbf{0.1451} & \textbf{0.1920} & \textbf{0.0948} & \textbf{0.1068} & \textbf{0.1370} & \textbf{0.1761} & \textbf{0.0921} & \textbf{0.1024} \\
\textcolor{red}{$\times$} & \textcolor{green!70!black}{$\checkmark$} & \textcolor{green!70!black}{$\checkmark$} & \textcolor{green!70!black}{$\checkmark$} & 0.0756 & 0.0993 & 0.0507 & 0.0551 & 0.0868 & 0.1132 & 0.0573 & 0.0641 & 0.0874 & 0.1067 & 0.0563 & 0.0639 \\
\textcolor{green!70!black}{$\checkmark$} & \textcolor{red}{$\times$} & \textcolor{green!70!black}{$\checkmark$} & \textcolor{green!70!black}{$\checkmark$} & 0.1049 & 0.1364 & 0.0699 & 0.0844 & 0.1243 & 0.1557 & 0.0838 & 0.0899 & 0.1197 & 0.1419 & 0.0774 & 0.0875 \\
\textcolor{green!70!black}{$\checkmark$} & \textcolor{green!70!black}{$\checkmark$} & \textcolor{red}{$\times$} & \textcolor{green!70!black}{$\checkmark$} & 0.0954 & 0.1143 & 0.0688 & 0.0701 & 0.1108 & 0.1451 & 0.0712 & 0.0845 & 0.1021 & 0.1308 & 0.0739 & 0.0782 \\
\textcolor{green!70!black}{$\checkmark$} & \textcolor{green!70!black}{$\checkmark$} & \textcolor{green!70!black}{$\checkmark$} & \textcolor{red}{$\times$} & 0.0677 & 0.0827 & 0.0468 & 0.0524 & 0.0798 & 0.1046 & 0.0532 & 0.0574 & 0.0717 & 0.0936 & 0.0491 & 0.0541 \\
\bottomrule
\end{tabular}
}
\end{table*}

\section{Experiments}
\label{section4}
In this section, we conduct extensive experiments on five widely used real-world datasets to validate the effectiveness of the proposed MAIL. Specifically, we compare MAIL with other SOTA MMRS methods under the same dataset settings and preprocessing protocols. In addition, we construct several variants of MAIL to examine the contribution of each component and verify whether each component effectively improves the final recommendation accuracy. Finally, we tune key hyperparameters to determine the optimal configuration. This study aims to answer the following research questions:
\begin{itemize}
    \item \textbf{RQ1:} How does MAIL perform compared with various SOTA models? (Section \ref{rq1})
    \item \textbf{RQ2:} How do different components in MAIL contribute to performance gains? (Section \ref{rq2})
    \item \textbf{RQ3:} Can MAIL construct more semantically aligned ID-free identity representations? (Section \ref{rq3})
        \item \textbf{RQ4:} How do different hyperparameter settings influence MAIL’s performance? (Section \ref{rq4})
    \item \textbf{RQ5:} How does MAIL perform compared with various SOTA models in different sparse data scenarios? (Section \ref{rq5})
       \item \textbf{RQ6:} Can MAIL mine more low-exposure semantic neighbors and alleviate popular-item-dominated propagation bias? (Section \ref{rq6})

\end{itemize}

\subsection{Experimental settings}
We conduct extensive experiments on five Amazon e-commerce datasets \cite{40-mcauley2015image, 10-xu2026survey}, including Baby, Sports, Clothing, Pet, and Beauty\footnote{{http://jmcauley.ucsd.edu/data/amazon/links.html}}\textsuperscript{,}\footnote{{https://drive.google.com/drive/folders/14mhuArJt-FwNEd-dSyvFD9Ey0ichdhGh}}, to evaluate the effectiveness of the proposed MAIL. The statistics of these datasets are reported in TABLE \ref{t-1}. Following most prior works \cite{3-mmrec, 15-li2025idfree, 29-ong2025spectrum}, we use 4096-dimensional visual features and 384-dimensional textual features, and adopt the 5-core setting to retain users and items. For each dataset, we randomly split historical interactions into training, validation, and test sets with a ratio of 8:1:1.

\subsubsection{Evaluation metrics}
Following existing MMRS studies \cite{3-mmrec, 11-yang2025fitmm, 10-xu2026survey}, we adopt two widely used metrics, Recall@\textit{K} and NDCG@\textit{K}, to evaluate model performance. Recall@\textit{K} measures how many relevant items in the test set appear in the Top-\textit{K} recommendation list, reflecting the model’s ability to retrieve items of interest. NDCG@\textit{K} further considers the ranking positions of relevant items and assigns higher weights to items ranked higher, measuring the ranking quality of recommendation results. In our experiments, we report the average Recall@\textit{K} and NDCG@\textit{K} over all test users with \textit{K}=10 and \textit{K}=20.

\subsubsection{Baseline}
To validate the effectiveness of MAIL, we benchmark it against several SOTA recommendation models. These baselines are grouped into three categories:

(1) Traditional recommendation models:
\begin{itemize}
    \item MF-BPR (UAI’09) \cite{26-rendle2009bpr}: This method models users’ pairwise preferences for positive items over negative items through triplets under implicit feedback.
    \item LightGCN (SIGIR’20) \cite{27-he2020lightgcn}: This method proposes a simplified graph convolutional recommendation framework that only retains neighborhood aggregation and layer-wise weighted fusion.
    \item DimCL (SIGKDD’25) \cite{41-zhang2025dimcl}: This method proposes a dimension-aware contrastive enhancement approach to identify and remove dimensional noise in augmented views.
    \end{itemize}
(2) ID-based multimodal recommendation models:
\begin{itemize}
    \item LGMRec (AAAI’24) \cite{28-guo2024lgmrec}: This method improves recommendation performance by decoupling local interests and modeling global dependencies.
    \item SMORE (WSDM'25) \cite{29-ong2025spectrum}: This method proposes a frequency-domain spectral fusion approach to suppress modality noise and improve multimodal recommendation performance.
    \item PGL (AAAI'25) \cite{42-yu2025mind}: This method proposes a main graph learning framework to improve recommendation performance by mining individual information from local structures.
    \item COHESION (SIGIR'25) \cite{43-xu2025cohesion}: This method proposes a framework that combines dual-stage fusion with composite graph convolution.
    \item SSR (NeurIPS'25) \cite{44-yang2026structured}: This method proposes a structured spectral reasoning framework that modulates fusion and alignment through frequency-band decomposition.
    \item SRGFormer (TMM'26) \cite{45-shi2026structurally}: This method proposes a structure-refined graph Transformer that improves recommendation performance by combining multi-head attention with hypergraph learning.
    \item LETTER (TMM'26) \cite{46-guo2025self}: This method improves recommendation performance by mining attribute combinations and adaptively adjusting user preference factors.
    \item SiMGR (AAAI'26) \cite{20-lian2026sign}: This method proposes a sign-aware multimodal graph recommendation framework that jointly models positive and negative feedback.
    \item DuGRec (INFFUS’26) \cite{39-ma2026let}: This method designs a self-supervised dual-graph reconstruction approach to denoise representations and alleviate the cold-start problem.
\end{itemize}

(3) ID-free multimodal recommendation models:
\begin{itemize}
    \item IDFREE (MM'25) \cite{15-li2025idfree}: This method proposes an ID-free multimodal collaborative filtering framework that replaces ID representations with multimodal features and positional encodings to improve recommendation performance.
\end{itemize}

\subsubsection{Implementation Details}
We implement the proposed MAIL model and all baseline models based on IDFREE \cite{15-li2025idfree} and MMRec \cite{3-mmrec}, respectively. All models are implemented based on PyTorch and evaluated on an RTX 3090 GPU.
Following a unified setting \cite{3-mmrec, 10-xu2026survey, 43-xu2025cohesion}, we fix the embedding dimension of users and items to 64 and initialize all parameters with Xavier \cite{47-glorot2010understanding} initialization. Moreover, we optimize all models using the Adam optimizer \cite{48-kingma2014adam}. During training, we set the batch size to 2048 and the maximum number of epochs to 1000. We adopt early stopping to prevent overfitting if Recall@20 on the validation set does not improve for 20 consecutive epochs.

\subsection{Performance Comparison (RQ1)} \label{rq1}
The detailed experimental results are reported in TABLE \ref{t-2}. The best results are highlighted in bold, and the second-best results are underlined. We have the following key observations:

(1) MAIL achieves the best results on Recall@10 and NDCG@10 across all five datasets. Compared with the strongest baseline IDFREE, MAIL improves Recall@10 by 13.78\%, 4.52\%, 5.03\%, 7.16\%, and 8.56\% on Baby, Sports, Clothing, Pet, and Beauty, respectively. The improvements on NDCG@10 are particularly significant on Baby and Beauty, indicating that MAIL not only retrieves more relevant items but also improves the ranking quality of recommendation lists. This is mainly because MAIL constructs content-aware ID-free identity representations through MAIC and mines low-exposure semantic neighbors through CSL, enhancing both semantic expressiveness and graph structure learning.

(2) Traditional recommendation models only use user-item interactions, making it difficult to fully represent item content and user preferences. Thus, their overall performance is relatively limited. ID-based multimodal recommendation methods introduce visual and textual features and usually outperform traditional collaborative filtering methods. This indicates that multimodal content can complement the semantic information missing from interaction signals. In contrast, IDFREE directly constructs user and item representations with multimodal features and positional encodings, outperforming most ID-based methods on all datasets. This shows that reducing the reliance on learnable ID embeddings and constructing representations from content semantics is effective in multimodal recommendation.

(3) Although IDFREE achieves strong performance, its static positional encodings still cannot dynamically adjust identity signals according to multimodal semantics. In addition, GNN-based multimodal recommendation methods are generally affected by popular-neighbor dominance, leading to insufficient mining of low-exposure semantic relations. To address these issues, MAIL introduces two modules, MAIC and CSL. MAIC modulates positional encodings through modality-aware gates to enhance the content awareness of identity representations. CSL mines low-exposure semantic neighbors via counterfactual neighbor augmentation and alleviates popular-item-dominated propagation bias. Therefore, MAIL further outperforms IDFREE on all five datasets, demonstrating that the proposed identity construction and structure learning strategies can effectively improve ID-free multimodal recommendation performance.

\begin{table}[t]
  \centering
  \caption{Generalization of MAIC on Different Backbones.}
  \label{t-4}
  \begin{tabular}{lcccc}
    \toprule
    \multirow{2}{*}{Dataset} & \multicolumn{2}{c}{Sports} & \multicolumn{2}{c}{Pet} \\
    \cmidrule(lr){2-3} \cmidrule(lr){4-5}
    & Recall@10 & NDCG@10 & Recall@10 & NDCG@10 \\
    \midrule
    PGL     & 0.0789 & 0.0428 & 0.1175 & 0.0644 \\
    +MAIC   & 0.0811 & 0.0442 & 0.1208 & 0.0658 \\
    \rowcolor{gray!20}
    Improvement & 2.79\% $\uparrow$ & 3.27\% $\uparrow$ & 2.81\% $\uparrow$ & 2.17\% $\uparrow$ \\
    \midrule
    DuGRec  & 0.0827 & 0.0462 & 0.1235 & 0.0696 \\
    +MAIC   & 0.0841 & 0.0475 & 0.1268 & 0.0718 \\
    \rowcolor{gray!20}
    Improvement & 1.69\% $\uparrow$ & 2.81\% $\uparrow$ & 2.67\% $\uparrow$ & 3.16\% $\uparrow$ \\
    \midrule
    IDFREE  & 0.1173 & 0.0770 & 0.1354 & 0.0876 \\
    +MAIC   & 0.1218 & 0.0803 & 0.1397 & 0.0912 \\
    \rowcolor{gray!20}
    Improvement & 3.84\% $\uparrow$ & 4.29\% $\uparrow$ & 3.17\% $\uparrow$ & 4.11\% $\uparrow$ \\
    \bottomrule
  \end{tabular}
\end{table}

\subsection{Ablation Study (RQ2)} \label{rq2}
In this subsection, we conduct comprehensive experiments to evaluate the effectiveness of each component in MAIL. Specifically, we disable key components of the proposed model to construct different variants and evaluate their performance. The results measured on three datasets are reported in TABLE \ref{t-3}. We denote the variants as follows:
\begin{itemize}
    \item \textbf{w/o MAIC}: We remove the Modality-Aware Identity Construction module, where modality-aware gates are no longer used to dynamically modulate positional encodings.
    
    \item \textbf{w/o CNA}: We remove the Counterfactual Neighbor Augmentation module, where counterfactual semantic neighbors are not constructed.
    
    \item \textbf{w/o SCE}: We remove the Structural Contrastive Enhancement module, where the cross-stage alignment and decoupled discrimination objectives are disabled.
    
    \item \textbf{w/o MM}: We remove multimodal information, where textual and visual modality features are not used for representation learning.
\end{itemize}

The experimental results show that removing any module leads to performance degradation. After removing MAIC, the model performance drops significantly, indicating that modality-aware identity construction enhances the content awareness of ID-free representations. Removing CNA also degrades performance, showing that counterfactual neighbor augmentation helps introduce low-exposure semantic neighbors and optimize the graph structure. Removing SCE makes the model less capable of preserving effective structural information, which demonstrates the necessity of structural contrastive enhancement for counterfactual structure learning. Removing MM causes the largest performance drop, indicating that multimodal information is the key foundation for identity construction and semantic structure learning in MAIL.

\begin{figure}[t]
    \centering    
    \includegraphics[width=\columnwidth]{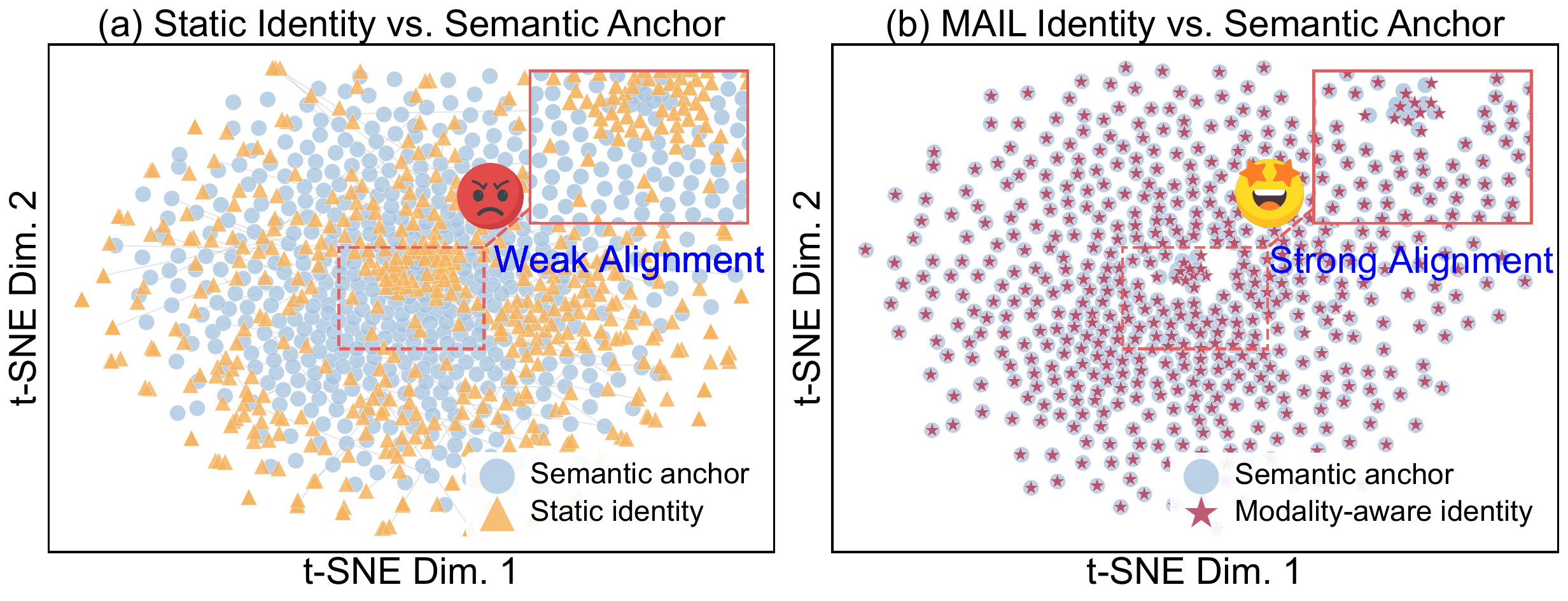}
    \caption{The t-SNE Visualization of identity-semantic alignment on the Baby dataset. }
    \label{f3}
\end{figure}

\begin{figure}[t]
    \centering    
    \includegraphics[width=\columnwidth]{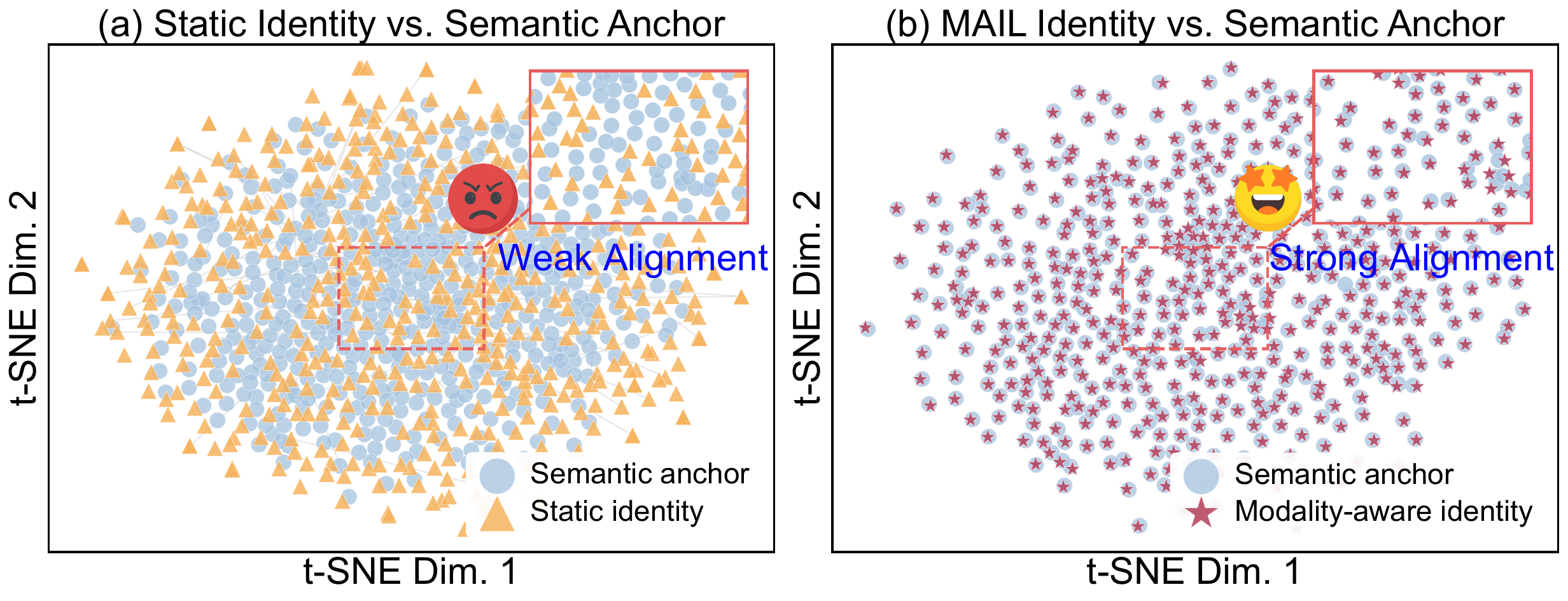}
    \caption{The t-SNE Visualization of identity-semantic alignment on the Sports dataset.}
    \label{f4}
\end{figure}

\begin{figure*}[t]
    \centering
     \includegraphics[width=1.0\textwidth]{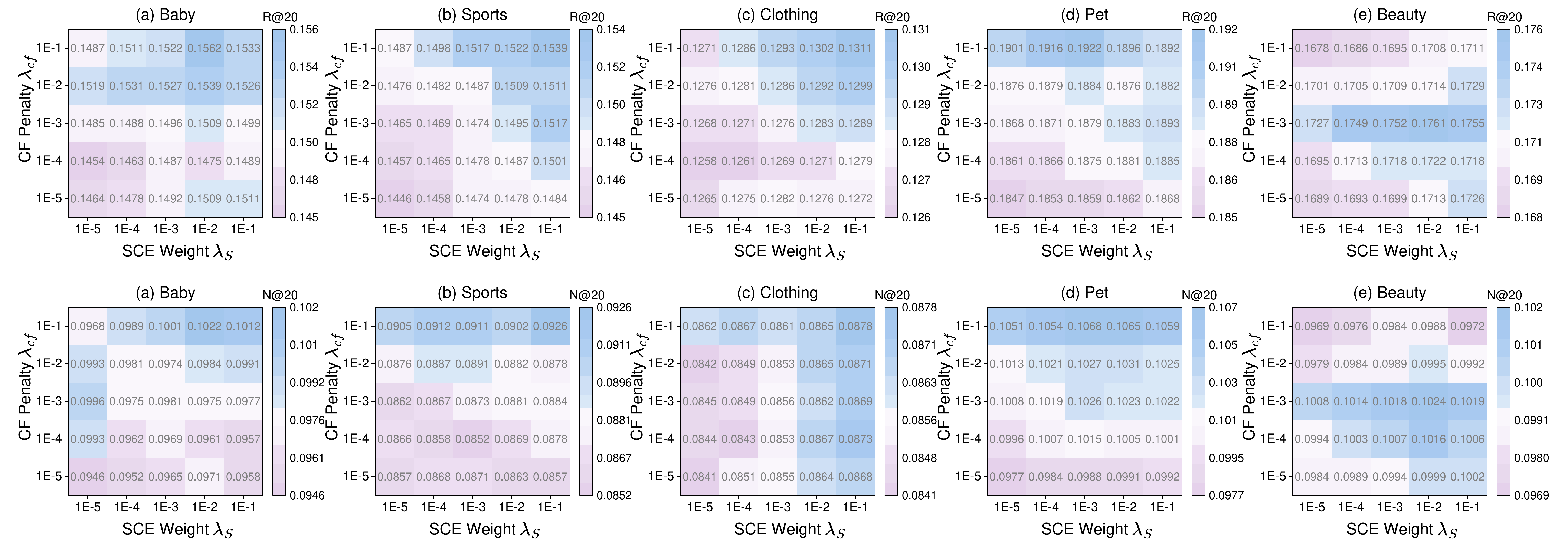}
    \caption{Effect of balancing hyperparameters $\lambda_{cf}$ and $\lambda_S$ on five datasets. 
$\lambda_{cf}$ controls the strength of popularity penalty in counterfactual neighbor selection, and $\lambda_S$ controls the weight of the structural contrastive enhancement loss $\mathcal{L}_{SCE}$. 
The upper and lower rows report Recall@20 and NDCG@20, respectively}
    \label{f7}
\end{figure*}

\begin{table}[t]
  \centering
  \small
  \caption{EFFECT OF MODALITY-AWARE FUSION WEIGHT $\alpha_p$.}
  \label{t-5}
  \resizebox{\columnwidth}{!}{
  \begin{tabular}{clcccccc}
    \toprule
    Dataset & Metrics & 0.3 & 0.4 & 0.5 & 0.6 & 0.7 & 0.8 \\
    \midrule
    \multirow{2}{*}{Baby} 
    & R@10 & 0.1276 & 0.1282 & 0.1294 & 0.1298 & \textbf{0.1305} & 0.1279 \\
    & N@10 & 0.0931 & 0.0937 & 0.0949 & 0.0943 & \textbf{0.0956} & 0.0947 \\
    \midrule
    \multirow{2}{*}{Sports} 
    & R@10 & 0.1198 & 0.1217 & \textbf{0.1226} & 0.1221 & 0.1213 & 0.1195 \\
    & N@10 & 0.0828 & 0.0835 & \textbf{0.0845} & 0.0841 & 0.0832 & 0.0822 \\
    \midrule
    \multirow{2}{*}{Clothing} 
    & R@10 & 0.1074 & 0.1087 & \textbf{0.1106} & 0.1101 & 0.1095 & 0.1077 \\
    & N@10 & 0.0798 & 0.0802 & \textbf{0.0828} & 0.0823 & 0.0808 & 0.0796 \\
    \midrule
    \multirow{2}{*}{Pet} 
    & R@10 & 0.1411 & 0.1429 & \textbf{0.1451} & 0.1443 & 0.1436 & 0.1414 \\
    & N@10 & 0.0923 & 0.0936 & \textbf{0.0948} & 0.0941 & 0.0932 & 0.0918 \\
    \midrule
    \multirow{2}{*}{Beauty} 
    & R@10 & 0.1347 & 0.1354 & 0.1366 & \textbf{0.1370} & 0.1363 & 0.1351 \\
    & N@10 & 0.0886 & 0.0907 & 0.0914 & \textbf{0.0921} & 0.0911 & 0.0897 \\
    \bottomrule
  \end{tabular}
  }
\end{table}

\subsection{Identity Alignment (RQ3)} \label{rq3}
We use t-SNE \cite{49-van2008visualizing} to visualize identity representations and multimodal semantic anchors. Specifically, we select item representations from the Baby and Sports datasets and project semantic anchor, static identity, and modality-aware identity into the same two-dimensional space. As shown in Fig. \ref{f3} and Fig. \ref{f4}, static identity shows a clear distribution shift from semantic anchor, while MAIL identity overlaps more closely with semantic anchor. This indicates that MAIL can construct more semantically consistent ID-free identity representations. We attribute this to the modality-aware gating mechanism in MAIC, which dynamically modulates positional encodings with textual and visual semantics and transforms static identity signals into content-aware identity representations.

In addition, we insert MAIC into three SOTA backbones, PGL, DuGRec, and IDFREE, to verify its generalization ability. As shown in TABLE \ref{t-4}, all three models achieve stable improvements on the Sports and Pet datasets after introducing MAIC. IDFREE obtains the most significant gains, with Recall@10 and NDCG@10 improved by up to 3.84\% and 4.29\%, respectively. We believe this is because IDFREE relies on positional encodings for ID-free identity construction and can better benefit from the modality-aware modulation mechanism of MAIC. These results show that MAIC is not only effective in MAIL but can also serve as a general identity enhancement module to improve the representation ability of multimodal recommendation models.

\subsection{Hyperparameter Analysis (RQ4)} \label{rq4}
We first analyze the effects of $\lambda_{cf}$ and $\lambda_s$ on MAIL. $\lambda_{cf}$ controls the popularity penalty strength in counterfactual neighbor selection, and $\lambda_s$ controls the weight of the structural contrastive enhancement loss $\mathcal{L}_{SCE}$. As shown in Fig. \ref{f7}, when both values are too small, the model cannot sufficiently mine low-exposure semantic neighbors. When they are too large, the model may overemphasize popularity penalization or structural constraints. Specifically, MAIL achieves better results on Baby when $\lambda_{cf}=10^{-1}$ and $\lambda_s=10^{-2}$, and performs best on Sports when $\lambda_{cf}=10^{-1}$ and $\lambda_s=10^{-1}$. These results show that moderate structural constraints and popularity penalization can better balance low-exposure semantic neighbor mining and recommendation optimization.

We further analyze the effect of the modality-aware fusion weight $\alpha_p$. $\alpha_p$ controls the fusion ratio between the textual gate $\gamma_x^t$ and the visual gate $\gamma_x^v$ in Modality-Aware PE. As shown in TABLE \ref{t-5}, the optimal $\alpha_p$ varies across datasets. Baby achieves the best result when $\alpha_p=0.7$, while Sports, Clothing, and Pet perform better when $\alpha_p=0.5$. This indicates that the contributions of textual and visual modalities vary across datasets, and an appropriate modality fusion ratio helps construct more effective content-aware ID-free identity representations.

\subsection{Different Sparse Data Scenarios (RQ5)} \label{rq5}
\begin{figure}[t]
    \centering    
    \includegraphics[width=\columnwidth]{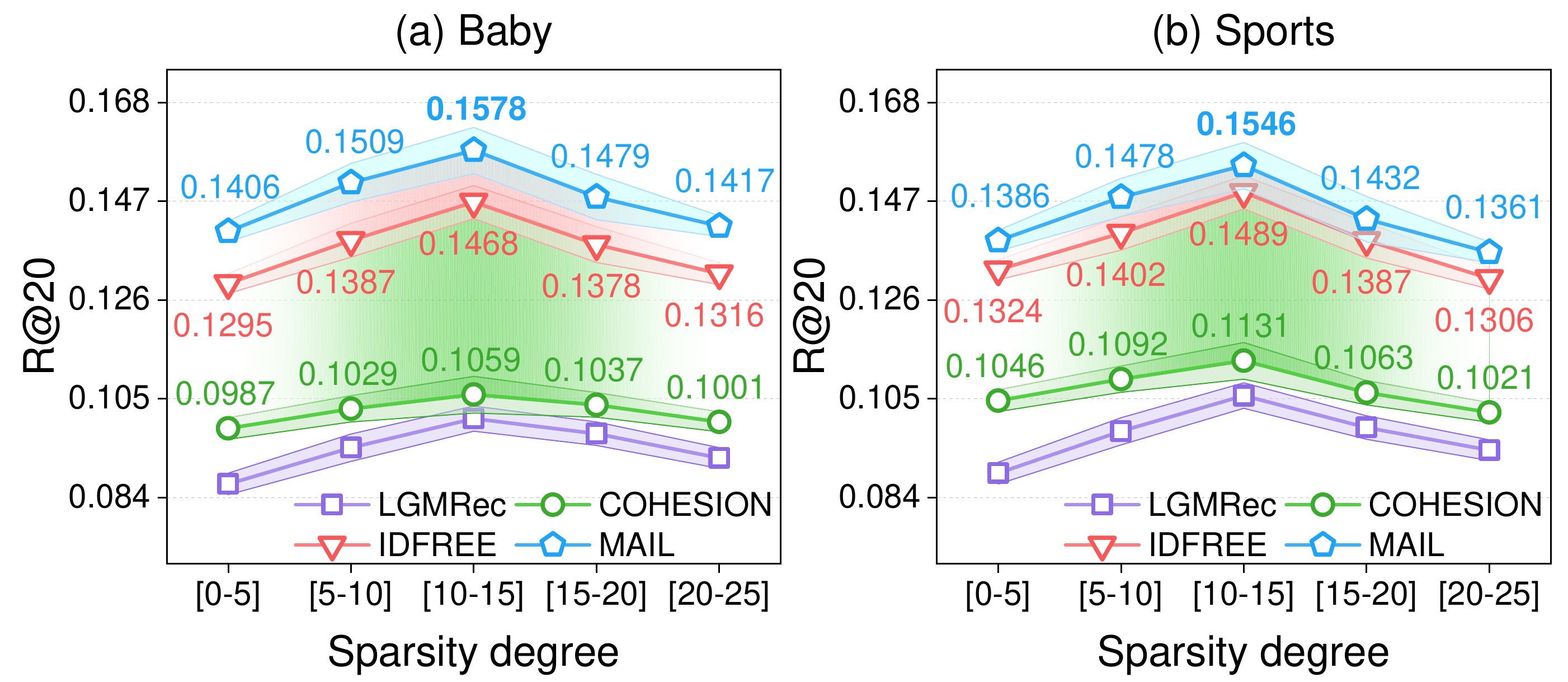}
    \caption{Sparsity-aware performance comparison on Baby and Sports datasets. Users are divided into different groups according to their historical interaction numbers, and Recall@20 is reported to evaluate model performance under different sparsity levels.}
    \label{f6}
\end{figure}

We further evaluate MAIL under different data sparsity scenarios. Specifically, we divide test users into several groups with different sparsity levels according to the number of interactions in the training set, and compare Recall@20 of MAIL with LGMRec, COHESION, and IDFREE on the Baby and Sports datasets. As shown in Fig. \ref{f6}, MAIL achieves superior performance across different sparse user groups and maintains stable improvements especially for users with few interactions. This result shows that MAIL can alleviate representation insufficiency caused by interaction sparsity through content-aware ID-free identity construction and counterfactual structure learning, improving recommendation robustness in sparse scenarios.

\subsection{Semantic Neighbors (RQ6)} \label{rq6}
\begin{figure}[t]
    \centering    
    \includegraphics[width=\columnwidth]{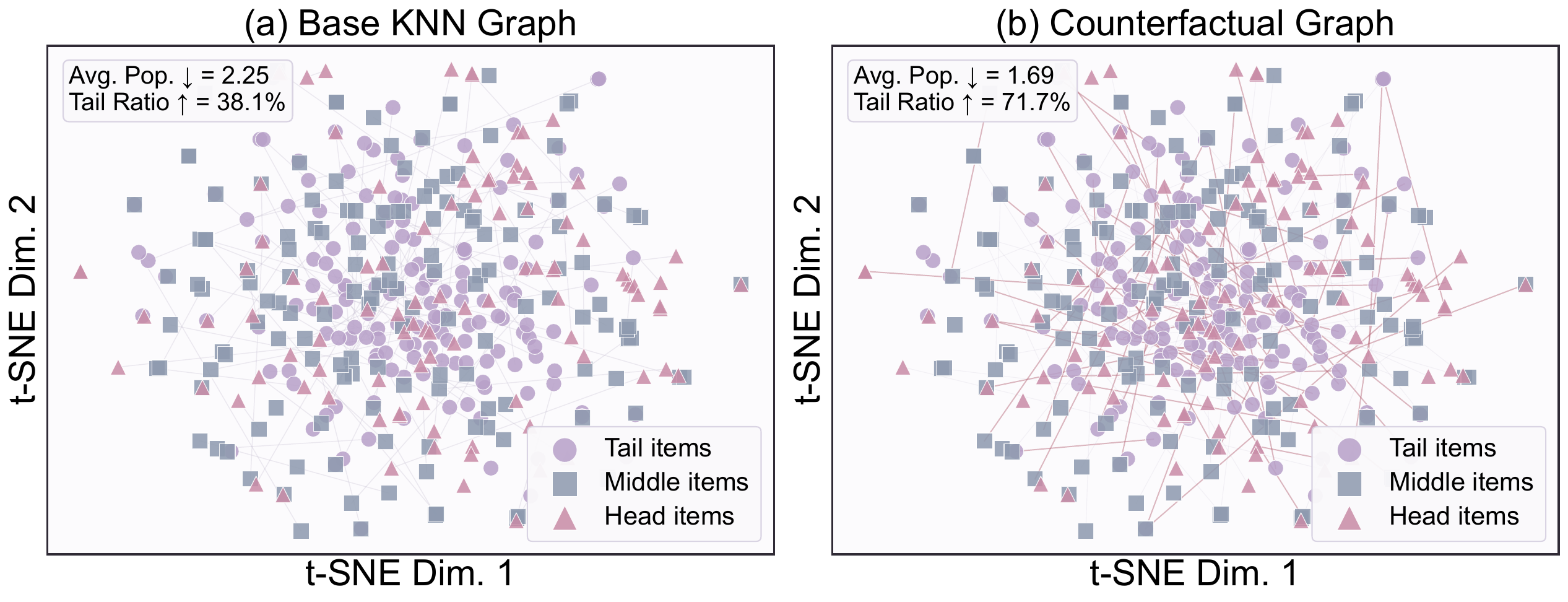}
    \caption{The t-SNE Visualization of counterfactual semantic neighbors on the Sports dataset. Avg. Pop. measures the average popularity of selected neighbors, where a lower value indicates less reliance on popular items. Tail Ratio measures the proportion of long-tail neighbors, where a higher value indicates better mining of low-exposure semantic neighbors.}
    \label{f5}
\end{figure}

To answer RQ6, we use t-SNE to project item representations into a two-dimensional space and visualize semantic neighbor connections in the base item-item KNN Graph and Counterfactual Graph. As shown in Fig. \ref{f5}, Avg. Pop. measures the average popularity of selected neighbors, where a lower value indicates less reliance on popular items. Tail Ratio measures the proportion of long-tail items among neighbors, where a higher value indicates that more low-exposure semantic neighbors are mined. The results show that the Counterfactual Graph reduces Avg. Pop. from 2.25 to 1.69 and increases Tail Ratio from 38.1\% to 71.7\%. This demonstrates that CSL can introduce more low-exposure yet semantically relevant neighbors through popularity penalty, thereby alleviating popular-item-dominated propagation bias.

\section{Conclusion}
\label{section5}
This paper analyzes the limitations of existing ID-free multimodal recommendation methods and further points out that static identity construction and popular-item-dominated graph learning limit semantic representation learning.
To address these issues, we propose a Modality-Aware Identity Construction and Counterfactual Structure Learning framework for ID-free multimodal recommendation, named MAIL.
Specifically, the modality-aware identity construction module dynamically modulates positional encodings with textual and visual semantics, transforming static identity signals into dynamic content-aware ID-free identity representations.
In addition, the counterfactual structure learning module mines low-exposure yet semantically relevant neighbors through popularity penalization, alleviating popular-item-dominated propagation bias.
Finally, we introduce structural contrastive enhancement to preserve counterfactual structural information during graph propagation and improve the discriminability of semantic representations.
Extensive experiments and ablation studies on five public datasets validate the effectiveness and generalization ability of MAIL.

\bibliographystyle{Tcyber/Bibliography/IEEEtran}
\bibliography{Tcyber/Bibliography/Bibfile}

\vspace{-1.2cm}
\begin{IEEEbiography}[{\includegraphics[width=1in, height=1.2in, clip, keepaspectratio]{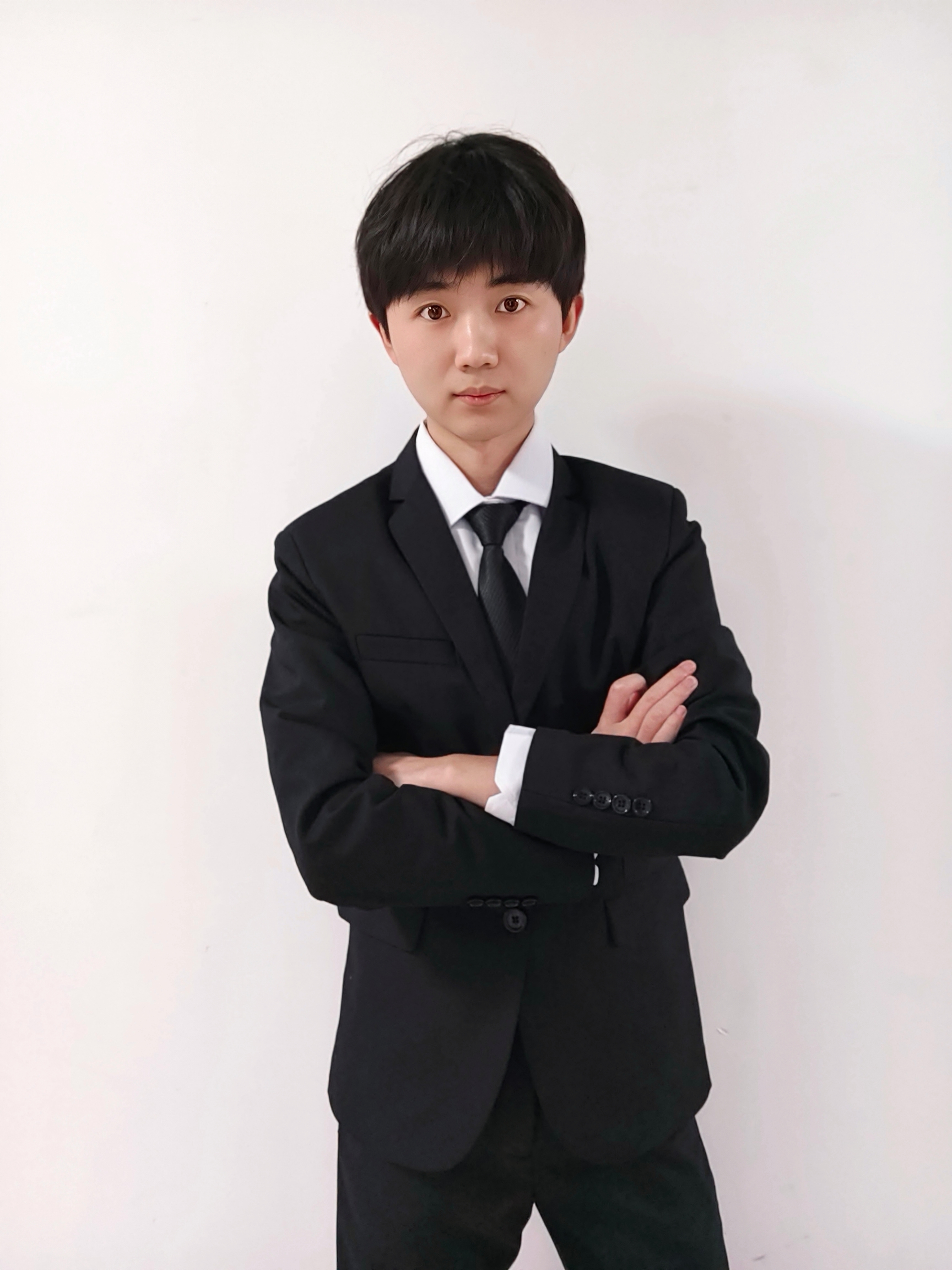}}] {Hongjian Ma} is currently pursuing the B.E. degree in Computer Science and Technology at Hubei University, Wuhan, China. His main research direction is recommendation systems.
\end{IEEEbiography}

\vspace{-1.2cm}
\begin{IEEEbiographynophoto}{Wenxin Huang} (Member, IEEE) received the B.S. degree in information technology from Metropolia University of Applied Sciences, Helsinki, Finland, in 2010, the M.S. degree in geoinformatics from Aalto University, Helsinki, in 2012, and the Ph.D. degree in computer science from Wuhan University, Wuhan, China, in 2020. She is currently a Lecturer with Hubei University, Wuhan, and a Research Officer with the Centre for Frontier AI Research, Agency for Science, Technology and Research (A*STAR), Singapore. Her research interests include multimedia content analysis and retrieval, large-scale multimedia data mining, and artificial intelligence.
\end{IEEEbiographynophoto}

\vspace{-1.2cm}
\begin{IEEEbiographynophoto}{Yan Zhang}
received the Ph.D. degree from Beihang University. He is currently a professor in the School of Computer Science at Hubei University. His research interests include cyberspace security and software engineering, among which he has in-depth theoretical research and practical experience in source code security detection and verification.
\end{IEEEbiographynophoto}

\vspace{-1.2cm}
\begin{IEEEbiography}[{\includegraphics[width=1in, height=1.2in, clip, keepaspectratio]{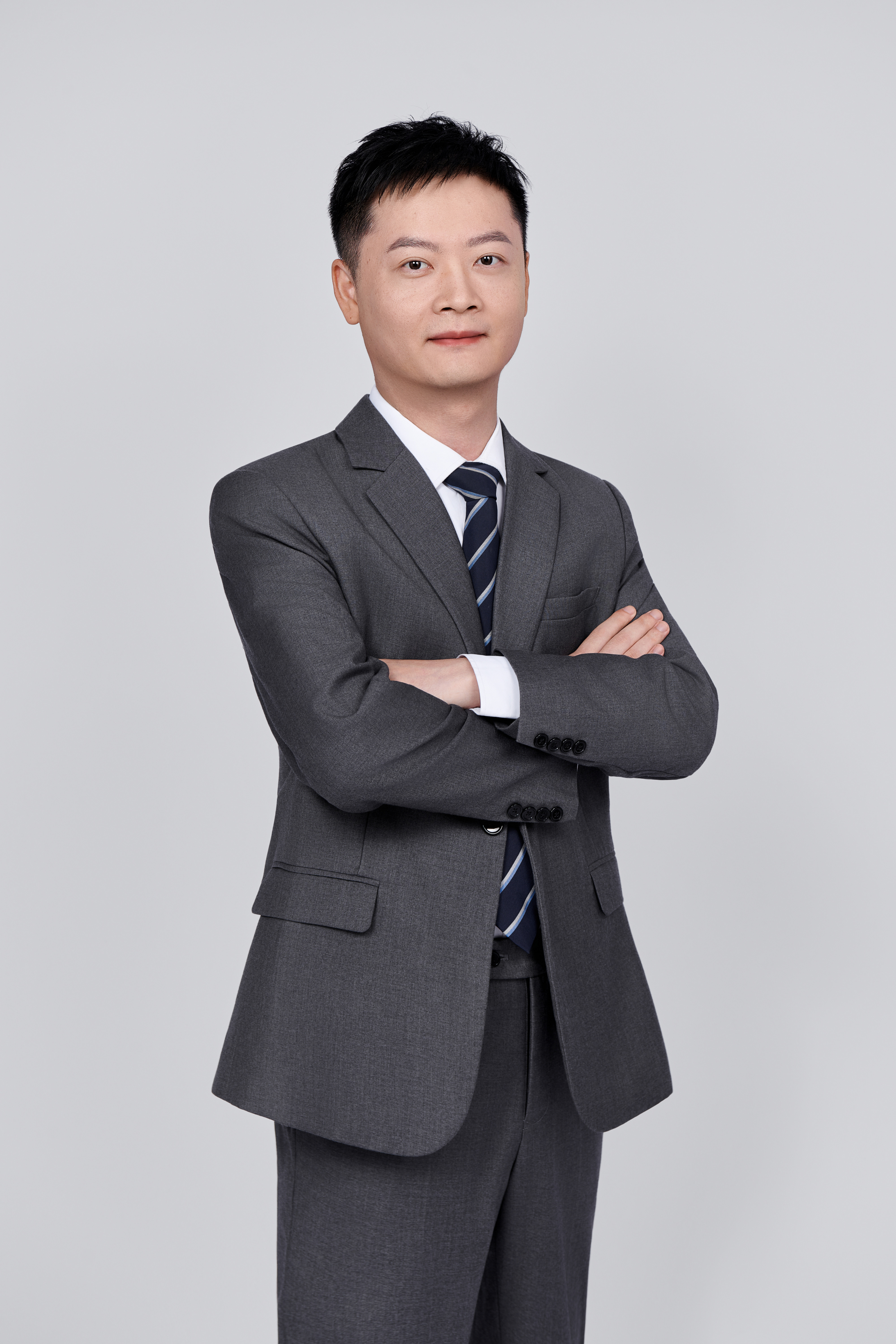}}] {Zhifei Li} (Senior Member, IEEE) received M.S. and Ph.D. degrees from the National Engineering Research Center for E-Learning at Central China Normal University in 2018 and 2021, respectively. Since October 2023, he has been an Associate Professor at the School of Computer Science, Hubei University. He has authored over forty peer-reviewed papers in international journals and conferences such as TKDE, TMM, SCIS, TNNLS, TKDD, AAAI, and IJCAI, with three papers selected as ESI highly cited papers. Additionally, he frequently reviews for international journals and conferences, including TKDE, IJCV, and AAAI. His research interests include knowledge graphs and recommendation systems.
\end{IEEEbiography}

\vspace{-1.2cm}
\begin{IEEEbiography}[{\includegraphics[width=1in, height=1.2in, clip, keepaspectratio]{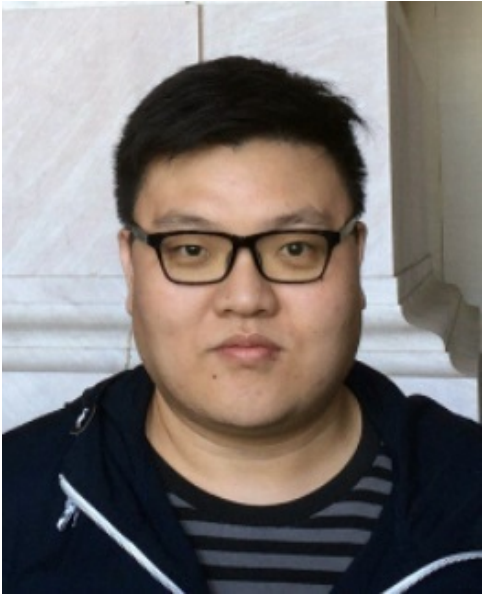}}] 
{Zheng Wang} (Senior Member, IEEE) received the B.S. and M.S. degrees from Wuhan University, China, in 2006 and 2008, respectively, and the Ph.D. degree from the National Engineering Research Center for Multimedia Software, School of Computer Science, Wuhan University, China, in 2017. From 2017 to 2020, he was a JST CREST Project Researcher and a JSPS Fellow at the National Institute of Informatics, Tokyo, Japan. From 2020 to 2021, he was a Project Assistant Professor at the Computer Vision and Media Laboratory, The University of Tokyo, Japan. He is a Professor in the School of Computer Science at Wuhan University. His research interests include multimedia content analysis and retrieval.
\end{IEEEbiography}

\end{document}